\documentclass[conference]{IEEEtran}
\IEEEoverridecommandlockouts
\usepackage{amsmath,amssymb}
\usepackage{amsthm}
\usepackage{graphicx}
\usepackage{hyperref}
\usepackage{booktabs}
\usepackage{algorithmic}
\usepackage{array}
\usepackage{multirow}
\usepackage[utf8]{inputenc}
\usepackage{cite}
\usepackage{color}
\usepackage{enumitem}
\usepackage{bm,soul}
\usepackage{algorithm}
\usepackage{placeins}

\newtheorem{theorem}{Theorem}

\theoremstyle{remark}

\theoremstyle{plain}
\newtheorem{thm}{Theorem}

\newtheorem{defi}[thm]{Definition}
\newtheorem{rem}{Remark}

\begin{document}
	
	\title{Cayley Graph Optimization for Scalable Multi-Agent Communication Topologies}

    \author{Jingkai~Luo and
        Yulin~Shao
\thanks{The authors are with the Department of Electrical and Computer Engineering, The University of Hong Kong, Hong Kong, China (e-mails: \url{lo563456@gmail.com}, \url{ylshao@hku.hk}). 
}
}

\maketitle

\begin{abstract}
Large-scale multi-agent communication has long faced a scalability bottleneck: fully connected networks require quadratic complexity, yet existing sparse topologies rely on hand-crafted rules. 
This paper treats the communication graph itself as a design variable and proposes CayleyTopo, a family of circulant Cayley graphs whose generator sets are optimized to minimize diameter, directly targeting worst-case information propagation speed.
To navigate the enormous search space of possible generator sets, we develop a lightweight reinforcement learning framework that injects a number-theoretic prior to favor structurally rich generators, alongside a message-propagation score that provides dense connectivity feedback during construction. 
The resulting CayleyTopo consistently outperforms existing hand-crafted topologies, achieving faster information dissemination, greater resilience to link failures, and lower communication load, all while approaching the theoretical Moore bound.
Our study opens the door to scalable, robust, and efficient communication foundations for future multi-agent systems, where the graph itself becomes optimizable rather than a fixed constraint.
\end{abstract}


\section{Introduction}
\label{sec:introduction}
Imagine a swarm of thousands of autonomous drones performing a coordinated search-and-rescue mission, or a fleet of connected vehicles navigating dense urban traffic. In such large-scale multi-agent systems, the ability to share information quickly and reliably is the bedrock of coordinated behavior \cite{chen2019uav,shao2024theory}. Yet, as the number of agents $N$ grows, the communication protocol that binds them faces a daunting challenge. A fully connected network, where every agent talks to every other, becomes an impossibility: its $\mathcal{O}(N^2)$ overhead would drown the system in congestion and scheduling conflicts. The central question is therefore how to design a sparse communication topology \cite{olfati2007consensus,li2025exponential,shao2021federated,das2019tarmac} that remains scalable, fast, and robust under realistic constraints.

The importance of communication structure has been recognized across multiple domains. In networked control theory, the connectivity of the graph directly influences the convergence rate of consensus \cite{olfati2007consensus}. In multi-agent reinforcement learning, differentiable communication schemes like CommNet \cite{sukhbaatar2016learning} and DIAL/RIAL \cite{foerster2016learning} enable agents to share information, and subsequent work has explored attentional or targeted communication \cite{das2019tarmac,jiang2018learning}. However, these methods primarily optimize message content and routing policies under a fixed or weakly constrained interaction pattern. The structure of the communication graph itself, i.e., which links exist, is rarely treated as a first-class design variable, especially under realistic degree constraints that reflect resource and hardware limits.

A recent advance made by Li et al.~\cite{li2025exponential} showed what is possible when structure is deliberately engineered. Their exponential topology-enabled scalable communication protocol (ExpoComm) employs a deterministic exponential graph, connecting each agent to peers at distances $2^0$, $2^1$, $\dots$. This rule-based topology achieves a logarithmic diameter $\lceil \log_2(N\!-\!1)\rceil$, and demonstrates strong scalability and zero-shot transferability on large benchmarks.

ExpoComm is a powerful example of how a fixed, hand-crafted topology can break the scaling bottleneck. Yet, it highlights a deeper opportunity: is the exponential rule itself optimal? This question motivates us to shift the perspective: \emph{instead of relying on hand-crafted rules, we treat the communication topology as an optimization variable}. Specifically, we model the network as a circulant Cayley graph \cite{godsil2013algebraic}, a family that generalizes the exponential construction and is parameterized by a set of generator steps. The design problem then becomes a discrete combinatorial optimization: find the generator set of limited size (i.e., respecting a degree budget) that minimizes the graph diameter, which directly governs worst-case information propagation speed.

Solving this optimization is non-trivial, as the search space of possible generator sets is enormous, and evaluating each candidate requires costly graph distance computations. To navigate this complexity, we develop a lightweight reinforcement learning (RL) framework tailored for topology selection. Our framework incorporates two key ideas: (1) a prior based on multiplicative order, which biases the search toward number-theoretically rich generators known to enhance connectivity; and (2) a message-propagation procedure that provides structural feedback during topology construction.

In summary, this paper makes the following contributions:
\begin{itemize}[leftmargin=0.5cm]
    \item We formulate multi-agent communication topology design as a discrete optimization over circulant Cayley graphs, directly targeting diameter minimization under degree constraints. The resulting family of topologies, {CayleyTopo}, achieves superior connectivity and propagation speed compared to existing hand-crafted constructions.
    \item To address the challenging combinatorial optimization, We develop a lightweight RL framework that integrates number-theoretic priors and a dense, propagation-based connectivity score. This enables efficient discovery of high-quality generator sets that respect the degree budget.
    \item Through comprehensive experiments, we demonstrate that CayleyTopo consistently outperforms rule-based baselines (ExpoComm, Fibonacci, primes) across multiple metrics: information dissemination latency, link-failure robustness, and communication load. Furthermore, our discovered topologies approach the theoretical Moore lower bound.
\end{itemize}

\section{System Model}
\label{sec:system}
	
We consider a multi-agent system consisting of $N$ agents, indexed by $0,1,\dots,N-1$. Communication occurs in discrete time steps. At each step, every agent may send messages to a subset of other agents, and the set of active communication links defines a graph. In this work, we focus on undirected graphs, meaning that if agent $i$ can send to agent $j$, then $j$ can also send to $i$.

\subsection{Communication Process and Key Metrics}
The communication structure is described by an undirected graph $\mathcal{G}=(\mathcal{V},\mathcal{E})$, where $\mathcal{V}=\{0,\dots,N-1\}$ and an edge $(i,j)\in\mathcal{E}$ indicates that agents $i$ and $j$ can exchange messages directly in a single time step. Messages can be relayed over multiple hops: if $i$ and $j$ are not directly connected, information can travel along a path $i = v_0$, $v_1$, $\dots$, $v_k = j$ where each consecutive pair is an edge. The number of hops needed to propagate a message from $i$ to $j$ is the length of the shortest such path, denoted $d(i,j)$.

Two fundamental metrics characterize the quality of a communication topology.

\begin{defi}[Diameter]
The diameter of $\mathcal{G}$ is the maximum shortest-path distance between any two agents: 
    \begin{equation}
        D(\mathcal{G})=\max_{u,v\in \mathcal{V}} d(u,v).
    \end{equation}
\end{defi}
The diameter represents the worst-case number of time steps required for a message to propagate from any agent to any other, assuming each step uses one-hop transmissions.

\begin{defi}[Degree]
For an undirected graph, the degree of agent $i$ is the number of its direct neighbors:
\begin{equation}
    \mathrm{deg}(i)=|\{j:(i,j)\in\mathcal{E}\}|.
\end{equation}
\end{defi}
The total number of edges $|\mathcal{E}|$ reflects the overall communication overhead. In practice, each agent has limited resources (e.g., bandwidth, power, or hardware constraints), which impose an upper bound on its degree. We denote this maximum allowed degree by $d_{\max}$, so the graph must satisfy $\deg(i) \le d_{\max}$ for all $i$.

The goal of topology design is to construct a graph with small diameter (for fast information propagation) while respecting a given degree budget $d_{\max}$.

\subsection{Exponential Graphs}

To make the above concepts concrete, we revisit the ExpoComm protocol proposed by Li et al.~\cite{li2025exponential}, which leverages exponential graphs as the underlying communication topology to achieve both scalability and low diameter.

\begin{figure}
    \centering
    \includegraphics[width=0.45\linewidth]{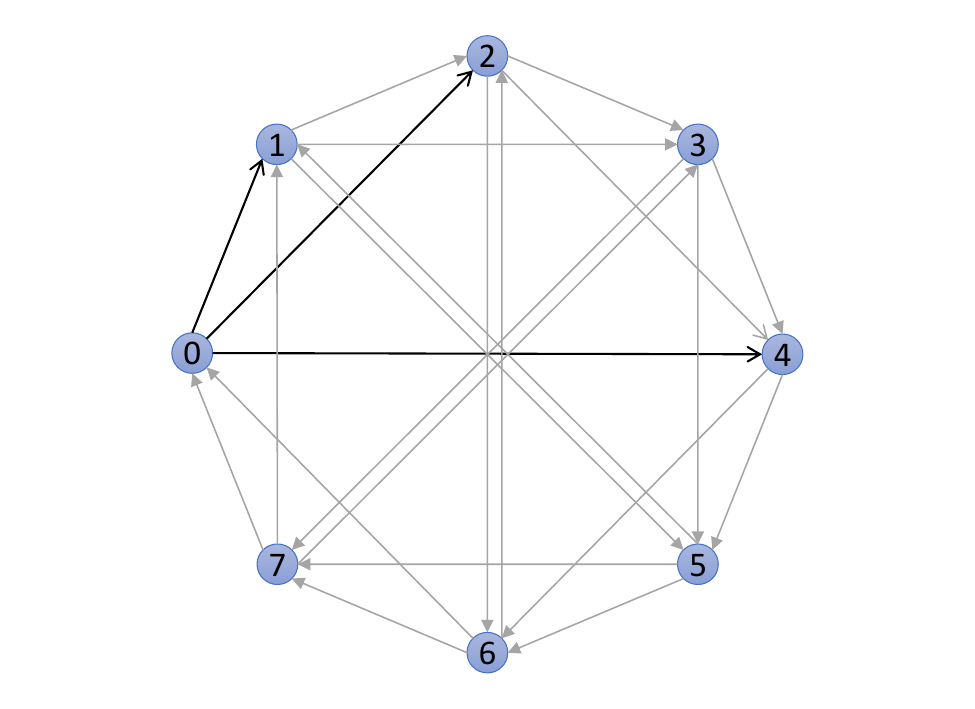}
    \caption{Illustration of the exponential graph for $N=8$ agents. Each agent connects to others at distances that are powers of two, resulting in a regular topology with diameter $\lceil \log_2 (N-1)\rceil = 3$.}
    \label{fig:expocomm_static}
\end{figure}

Exponential graphs are defined on the cyclic ordering of agents, as shown in Fig.~\ref{fig:expocomm_static}, and follow a deterministic rule: each agent $i$ connects to agents at distances that are powers of two, i.e.,
\begin{equation}
    i \pm 2^0, i \pm 2^1, \dots, i \pm 2^{\lceil \log_2(N-1)\rceil}~(\mathrm{mod}~N).
\end{equation}
This yields a regular graph where 
\begin{itemize}[leftmargin=0.32cm]
    \item Every agent has degree $2\lceil \log_2(N-1)\rceil$ (or $\lceil \log_2(N-1)\rceil$ if only outgoing edges are counted in a directed interpretation).
    \item Its diameter is $\lceil \log_2(N-1)\rceil$: any integer distance $d$ between $0$ and $N-1$ can be written as a sum of distinct powers of two (its binary representation). A message can therefore reach an agent at distance $d$ by taking jumps corresponding to the bits of $d$, using at most $\lceil \log_2(N-1)\rceil$ hops.
\end{itemize} 

The exponential graph family thus demonstrates that a carefully chosen communication topology can achieve a logarithmic diameter with degree (or per-step edge count) that grows only logarithmically in $N$, making it highly scalable.
	
\subsection{Generalizing the Topology Design Problem}\label{sec:IIB}

In the exponential graph, each agent $i$ connects to agents at offsets that are powers of two. Critically, the connection pattern is identical for every agent: the neighborhood of any agent looks the same up to a cyclic shift of indices. This property, known as {vertex-transitivity}, is not unique to the exponential construction. It suggests that we can capture a broad family of structured topologies using Cayley graphs on the cyclic group. 

\begin{defi}[Cayley graph on $\mathbb{Z}_N$]
Let $\mathbb{Z}_N = \{0,\allowbreak 1,\allowbreak \dots,\allowbreak N-1\}$ denote the additive group of integers modulo $N$. A Cayley graph on this group is defined by a set $S \subset \mathbb{Z}_N \setminus {0}$, called the {generator set}. The resulting graph $\operatorname{Cay}(\mathbb{Z}_N, S)$ has vertex set $\mathbb{Z}_N$, and for every vertex $u$ and every generator $s \in S$, there is a directed edge from $u$ to $u+s$ (mod $N$).
\end{defi}

To obtain an undirected graph as in our multi-agent setting, we impose that $S$ is symmetric: if $s \in S$ then its additive inverse $-s$ (mod $N$) must also be in $S$. In that case, each generator $s$ contributes two undirected edges per vertex: one to $u+s$ and one to $u-s$. To avoid redundancy, we often describe the graph using a set of {positive generators} $S^+ \subset \{1,\dots,N-1\}$ and then set $S = S^+ \cup (-S^+)$. The degree of every vertex is then $|S| = 2|S^+|$. Each positive generator $s \in S^+$ acts as a fixed step size: every agent $i$ can communicate directly with agents $i+s$ and $i-s$ (mod $N$). The entire graph is built by applying these steps from every vertex.

The exponential graph from ExpoComm is a special case of this construction. Its positive generator set is precisely the powers of two: $S^+_{\mathrm{exp}}= \{2^0, 2^1, \dots, 2^{\lceil \log_2(N-1)\rceil}\}$.

Instead of fixing $S^+$ to the standard ExpoComm (powers-of-two) generator set, this paper treats it as a design variable. The goal is to select a set of positive generators, subject to a degree budget, that yields a graph with favorable communication properties. Since the graph is vertex-transitive, its global characteristics, such as the worst-case number of hops needed for a message to travel between any two agents, are determined solely by the generator set.

Specifically, we denote by $D(S^+)$ the diameter of $\operatorname{Cay}(\mathbb{Z}_N, S)$. For a given degree budget $d_{\max}$ (the maximum number of neighbors each agent can support), the undirected degree of each vertex is $2|S^+|$, so we must satisfy $2|S^+| \le d_{\max}$, i.e., $|S^+| \le \lfloor d_{\max}/2 \rfloor$. Let $K \triangleq \lfloor d_{\max}/2 \rfloor$ denote the maximum number of positive generators.

Our core optimization problem then becomes:
\begin{equation}
\min_{S^+ \subset \mathcal{C}} D(S^+), \quad \text{s.t.} \quad |S^+| \le K,
\label{eq:optimization}
\end{equation}
where $\mathcal{C}$ is a candidate pool of positive integers. To ensure the graph is connected, the set $S^+$ must be able to generate the whole group $\mathbb{Z}_N$, which holds if and only if the greatest common divisor (gcd) of all elements in $S$ and $N$ is $1$. A convenient sufficient condition is to restrict all generators to be coprime to $N$; then any non-empty subset automatically satisfies this gcd condition. Thus, we restrict $\mathcal{C}$ to numbers that are coprime to $N$ (e.g., all such numbers, or a subset like primes less than $N$).

\begin{rem}
The diameter $D(S^+)$ serves as the primary objective, as it directly captures worst-case information dissemination latency. In practice, when two different generator sets yield the same diameter, we may compare them using the average shortest-path length
\begin{equation}
    L(S^+) = \frac{1}{N(N-1)}\sum_{u\neq v} d_{S^+}(u,v),
\end{equation}
where $d_{S^+}(u,v)$ is the distance in the corresponding Cayley graph. $L(S^+)$ is useful as a tie-breaker when observed diameters are small and many candidates achieve the same $D(S^+)$.
\end{rem}

By treating topology design as a discrete optimization over generator sets, we open the door to systematically searching for structures that go beyond fixed designs such as ExpoComm. However, the search space of possible generator sets is enormous: the number of subsets of $\mathcal{C}$ grows combinatorially, making exhaustive enumeration infeasible. Moreover, evaluating $D(S^+)$ requires computing all-pairs shortest paths (or at least distances from a single source due to vertex-transitivity), which becomes expensive as $N$ scales. Therefore, an efficient optimization method is required.
	
\section{Topology Optimization via RL with Number-Theoretic Priors}
\label{sec:methods}
	
In this section, we present a RL framework that integrates number-theoretic priors and a structural connectivity score to discover near-optimal Cayley graph topologies under a degree budget.

\subsection{Theoretical Baseline and The Multiplicative Order}
Before searching for optimal topologies, it is useful to understand the fundamental limits of what a graph can achieve. For a regular graph where every vertex has degree $\Delta$, the number of vertices reachable within $D$ hops is bounded below.
\begin{theorem}[Moore bound \cite{miller2012moore}]
        For a $\Delta$-regular graph with diameter $D$, the total number of vertices $N$ cannot exceed
	\begin{equation}\label{eq:moore}
		N \le 1 + \Delta \sum_{h=0}^{D-1} (\Delta-1)^h.
	\end{equation}
\end{theorem}
This bound arises from a breadth-first expansion argument: starting from a single vertex, at most $\Delta$ new vertices can be reached in the first step, and at most $(\Delta-1)$ new vertices from each subsequent step, as one neighbor is already accounted for. For a given $N$ and degree $\Delta$, this inequality provides a lower bound on the achievable diameter $D$. In our context, $\Delta = 2K$, the undirected degree of the circulant graph. The Moore bound serves as a theoretical reference: no topology can have a diameter smaller than the smallest $D$ satisfying \eqref{eq:moore}. We will use this bound to gauge how close our discovered topologies come to the theoretical optimum.

The search space for generator sets is enormous, but not all generators are equally promising. A classical concept from number theory offers a powerful heuristic for identifying generators that promote good connectivity.

\begin{defi}[Multiplicative order \cite{ireland1990classical}]
For an integer $a$ coprime to $N$, the multiplicative order of $a$ modulo $N$, denoted $\operatorname{ord}_N(a)$, is the smallest positive integer $k$ such that $a^k \equiv 1 \pmod{N}$.
\end{defi}

This quantity is directly tied to the structure of the multiplicative group $\mathbb{Z}_N^\times$. A generator $a$ with a large multiplicative order generates a large cyclic subgroup of $\mathbb{Z}_N^\times$. The residues $a^0, a^1, \dots, a^{k-1}$ are therefore spread relatively uniformly across the set of numbers coprime to $N$. When these residues are used as additive step sizes in a circulant graph, they tend to cover the additive group $\mathbb{Z}_N$ more evenly, which can lead to smaller diameters and average path lengths. While this relationship is heuristic for composite $N$, empirical evidence and insights from additive number theory \cite{tao2006additive} suggest that high-order generators are strong candidates for constructing low-diameter graphs.

To incorporate this prior into our optimization, we precompute for each candidate generator $p \in \mathcal{C}$ its normalized multiplicative order:
\begin{equation}\label{eq:order}
    \omega(p) = \frac{\operatorname{ord}_N(p)}{\max_{q \in \mathcal{C}} \operatorname{ord}_N(q)}.
\end{equation}
This normalization maps the order to the interval $[0,1]$, allowing us to use it as a scalar feature and a bias term in our RL agent. The prior effectively guides the search toward generators that are number-theoretically rich, reducing the need to explore less promising candidates.

\subsection{RL for Generator Selection}
\label{subsec:rl}

We now transform the topology design problem as a sequential decision process. The goal is to construct a generator set $S^+$ of size $K$ by selecting $K$ generators from the candidate pool $\mathcal{C}$ without replacement. This process is naturally modeled as an RL episode with $K$ steps.

\subsubsection{State and action spaces}
At step $t$, where $t = 1, \dots, K$, the state encodes the history of selections and the current progress. It consists of:
\begin{itemize}[leftmargin=0.5cm]
\item The set of already selected generators $\mathcal{I}_{\text{sel}}^{t-1}$, which implicitly defines the partial positive generator set $S^+_{t-1}$.
\item A normalized step index $t/K$, indicating how many selections remain.
\item For each candidate generator $p_i \in \mathcal{C}$, its normalized multiplicative order $\omega(p_i)$ and its index $i/|\mathcal{C}|$ (a positional feature).
\end{itemize}

The action space is the set of candidates not yet selected. The policy $\pi(a_t \mid \mathrm{state})$ is a probability distribution over these available actions.

\subsubsection{Policy network}
We use a lightweight multilayer perceptron (MLP) to parameterize the policy. For each candidate $p_i$, the input feature vector is $\bm{x}_{t,i} = \left[\omega(p_i), {i}/{|\mathcal{C}|},{t}/{K}\right]^\top$.
These features are passed through a two-layer MLP:
\begin{equation}
    s_{t,i} = \bm{W}_2 \tanh(\bm{W}_1 \bm{x}_{t,i} + \bm{b}_1) + \bm{b}_2 + \eta \cdot \omega(p_i),
\end{equation}
where $\bm{W}_1$, $\bm{W}_2$, $\bm{b}_1$, $\bm{b}_2$ are learnable parameters of the MLP; $\eta$ is a bias coefficient that amplifies the influence of the multiplicative order prior. The logits are then masked to exclude already chosen generators, and a softmax yields the policy
\begin{equation}
    \pi_t(i) = \frac{\exp(s_{t,i}) \cdot \mathbb{I}_{\{i \notin \mathcal{I}_{\text{sel}}\}}}{\sum_{j \notin \mathcal{I}_{\text{sel}}} \exp(s_{t,j})}.
\end{equation}
Note that the order prior is injected both implicitly via input features and explicitly via the additive logit bias term $\eta\cdot\omega(p_i)$, improving exploration stability in early training.

\subsubsection{Reward}
The ultimate objective is to minimize the diameter $D(S^+_K)$. However, providing this as a sparse reward only at the end of the episode makes learning difficult. We therefore augment it with a dense, step-wise reward signal that reflects the quality of the partial topology. This signal has two components.

First, we track the average normalized multiplicative order of the selected generators:
\begin{equation}\label{eq:averageorder}
\omega_t = \frac{1}{t}\sum_{j=1}^{t} \omega(p_{a_j}).
\end{equation}
Second, we compute a structural score $g(S^+_t)$ based on a message-propagation procedure, which will be detailed in the next subsection. This score provides immediate feedback on how well information can spread through the partial graph.

The step-wise shaped reward is then defined as the incremental improvement in these metrics:
\begin{equation}\label{eq:densereward}
r_t^{\text{shape}} = \lambda (\omega_t - \omega_{t-1}) + \lambda_g (g(S^+_t) - g(S^+_{t-1})),
\end{equation}
with $\omega_0 = g(S^+_0) = 0$. At the final step $t=K$, we compute the exact diameter $D(S^+_K)$ (via breadth-first search from vertex $0$, exploiting vertex-transitivity) and incorporate it into the reward:
\begin{equation}\label{eq:reward_final}
r_K = r_K^{\text{shape}} - D(S^+_K).
\end{equation}
The total return for an episode is therefore
\begin{equation}
R = \sum_{t=1}^{K} r_t = -D(S^+_K) + \lambda \omega_K + \lambda_g g(S^+_K).
\label{eq:reward}
\end{equation}
This formulation allows the agent to receive dense feedback during topology construction while ensuring that minimizing the diameter remains the primary objective of the return.

Our complete framework operates as follows. The RL agent sequentially selects $K$ generators from a candidate pool $\mathcal{C}$. At each step, its policy is influenced by a prior that favors numbers with high multiplicative order \eqref{eq:order}. The agent receives dense rewards \eqref{eq:densereward}, based on the average order of selected generators \eqref{eq:averageorder} and a structural score \eqref{eq:score} derived from a fast message propagation (Section \ref{sec:connectivity-score}). At the end of each episode, the exact diameter of the final topology is computed and incorporated into the final reward \eqref{eq:reward_final}. 

The policy is trained using proximal policy optimization (PPO) \cite{schulman2017ppo}. Our implementation follows the standard PPO but is adapted for our problem. Key aspects include:
\begin{itemize}[leftmargin=0.5cm]
\item We collect a batch of complete episodes (each constructing a topology of $K$ generators) before performing updates.
\item A separate MLP, with inputs $[\omega_t, g(S^+_t)]^\top$, estimates the value function of the state. This low-dimensional input focuses the value network on the key metrics of progress.
\item We compute generalized advantage estimation (GAE) \cite{schulman2015gae} using the step rewards and value predictions.
\item For each candidate topology, we cache computed graph metrics (diameter, average path length, etc.) to avoid redundant computation.
\end{itemize}

\subsection{Connectivity Scoring via Message Propagation}
\label{sec:connectivity-score}

The structural score $g(S)$ used in the reward function is designed to provide a smooth, differentiable proxy for the graph's communication efficiency. It is derived from a fixed, parameter-free message-propagation procedure that simulates how quickly local information spreads through the graph.

Given a generator set $S$, we first construct its undirected adjacency matrix $\bm{A} \in \mathbb{R}^{N \times N}$, where $\bm{A}_{i,j}=1$ if $j = i \pm s \pmod{N}$ for any $s \in S$. We then add self-loops to obtain $\widetilde{\bm{A}} = \bm{A} + \bm{I}$. The normalized propagation operator is
\begin{equation}
\hat{\bm{A}} = \widetilde{\bm{D}}^{-1/2} \widetilde{\bm{A}} \widetilde{\bm{D}}^{-1/2},
\end{equation}
where $\widetilde{\bm{D}}$ is the diagonal degree matrix of $\widetilde{\bm{A}}$.

Each node $i$ is initialized with a fixed three-dimensional feature:
\begin{equation}
    \bm{x}_i^{(0)} =
    \left[\cos\!\left(\frac{2\pi i}{N}\right),\
    \sin\!\left(\frac{2\pi i}{N}\right),\
    1\right].
\end{equation}
Stacking all node features yields $\bm{X}^{(0)}\in\mathbb{R}^{N\times 3}$.
We apply two rounds of nonlinear propagation:
\begin{equation}
    \bm{X}^{(k+1)} = \tanh\!\left(\hat{\bm{A}}\bm{X}^{(k)}\right),\quad k=0,1.
\end{equation}
This operation measures how quickly local information mixes over the graph.

After two propagation steps, we compute the variance $v_j$ of each feature channel $j$ across all nodes: $v_j=\mathrm{Var}\!\left(\bm{X}^{(2)}_{:,j}\right)$. The structural score is the negative sum of these variances:
\begin{equation}\label{eq:score}
    g(S) = -\sum_{j=1}^{3} v_j.
\end{equation}
A graph that mixes information quickly will have node features that become similar after a few propagation steps, leading to low variance and thus a high score $g(S)$. Conversely, a poorly connected graph will retain high variance. This procedure is computationally lightweight (relying on sparse matrix operations) and provides a dense, deterministic measure of global connectivity that can be used to shape the reward at every step of the RL process.

\section{Experimental Results and Discussion}
\label{sec:results}

This section evaluates the communication topologies discovered by our RL-based optimization framework. All experiments are conducted with $N=1024$ agents, and each topology is constrained to an undirected degree of $14$ (i.e., $K=7$ positive generators). We compare our method, denoted by {CayleyTopo}, against several baselines:
\begin{itemize}[leftmargin=0.5cm]
\item \textit{ExpoComm}: the exponential-graph baseline from Li et al.~\cite{li2025exponential} (generators $2^0,2^1,\ldots$).
\item \textit{Fibonacci}: generators taken from the Fibonacci sequence.
\item \textit{Simple Prime}: the first $K$ primes in increasing order.
\item \textit{Broadcast}: a fully connected network (or a broadcast protocol) that serves as a high-overhead baseline.
\end{itemize}

All topologies are evaluated under the same degree budget, ensuring a fair comparison of their communication efficiency and robustness.

\subsection{Information Dissemination Latency}

We first examine how quickly a message can spread through the network. We simulate a push-style epidemic process: at each round, every informed node attempts to transmit a message to each of its neighbors, with a per-link success probability of $0.75$. The process continues until all nodes are informed, or until a maximum of $120$ rounds is reached. For each topology, we run $30$ independent trials and report the average number of rounds needed to reach $90\%$ ($T90$) and $100\%$ ($T100$) of the agents, as well as the total number of transmissions (AvgTX).

\begin{figure}[!t]
	\centering
	\includegraphics[width=0.38\textwidth]{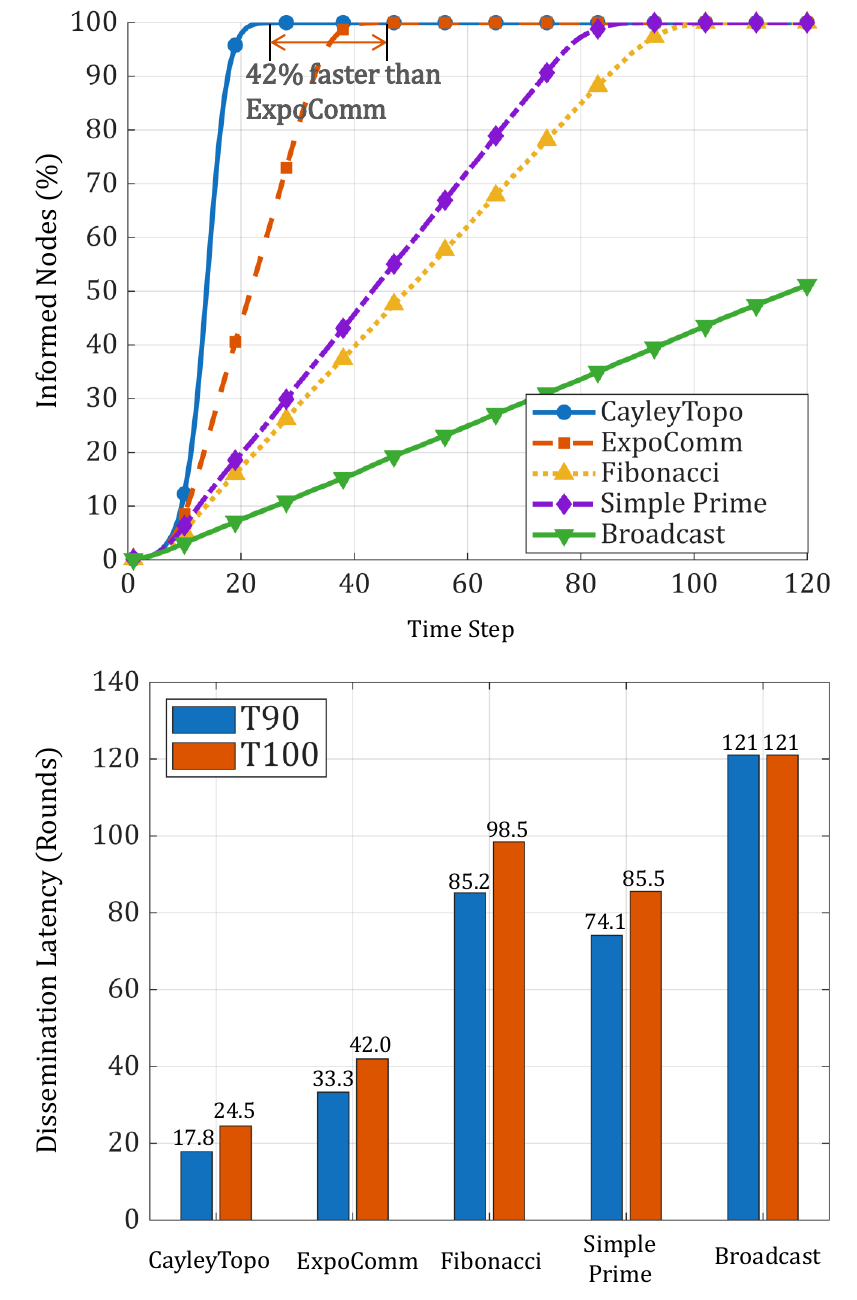}
	\caption{Information dissemination latency under degree budget $14$: average rounds to $90\%$ ($T90$) and $100\%$ ($T100$) informed agents.}
	\label{fig:dissemination-latency}
\end{figure}

Fig.~\ref{fig:dissemination-latency} compares mean $T90$ and $T100$ across topologies. CayleyTopo achieves the fastest dissemination, with $T90\approx17.8$ rounds and $T100\approx24.5$. {ExpoComm} requires about $33.3$ rounds for $90\%$ coverage and about $42.0$ for full coverage, while Fibonacci and Simple Prime need roughly $74$-$98$ rounds depending on the threshold. Broadcast hits the $120$-round cap without full coverage. The AvgTX of CayleyTopo is about $1.05\times10^4$, about $2.02\times10^4$ for ExpoComm, about $3.05\times10^4$ for broadcast, about $4.28\times10^4$ for Simple Prime, and about $4.98\times10^4$ for Fibonacci. Thus, CayleyTopo attains both the lowest latency and the lowest AvgTX among the designs shown, improving over ExpoComm on both metrics.
	
\subsection{Link-Failure Robustness}

\begin{figure}[!t]
    \centering
    \includegraphics[width=0.42\textwidth]{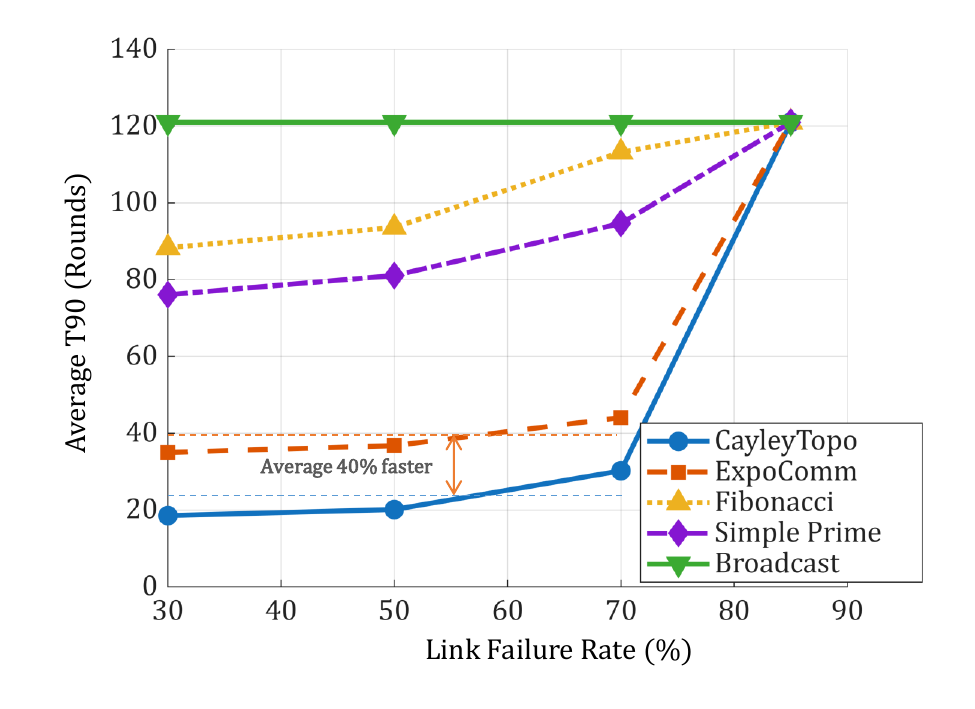}
    \caption{Average dissemination delay ($T90$, rounds) versus link failure rate.}
    \label{fig:link-failure-robustness}
\end{figure}

A practical communication topology must tolerate link failures. We evaluate robustness by randomly removing edges at rates of $30\%$, $50\%$, $70\%$, and $85\%$. For each failure rate, we generate $20$ random failure realizations per topology. We report (i) the average rounds to $90\%$ informed nodes ($T90$) under random failures, illustrated in Fig.~\ref{fig:link-failure-robustness}, and (ii) at $85\%$ failures, the size of the largest connected component (LCC) as a percentage of agents, together with $\Pr[\mathrm{LCC}\ge80\%]$ under random (Pr80-R) and distance-biased (Pr80-D) removals, summarized in Table~\ref{tab:robustness-85}.

\begin{table}[!t]
    \centering
    \small
    \setlength{\tabcolsep}{4pt}
    \renewcommand{\arraystretch}{1.15}
    \caption{Robustness at $85\%$ link failures: mean LCC size (\% of agents) and $\Pr[\mathrm{LCC}\!\ge\!80\%]$ under random (Pr80-R) and distance-biased (Pr80-D) removals.}
    \label{tab:robustness-85}
    \begin{tabular*}{\columnwidth}{@{\extracolsep{\fill}}@{}l r r r@{}}
        \toprule
        Topology & LCC@85\% & Pr80-R & Pr80-D \\
        \midrule
        CayleyTopo & 84.27\% & 100\% & 100\% \\
        ExpoComm & 83.12\% & 85\% & 0\% \\
        Fibonacci & 69.22\% & 55\% & 0\% \\
        Simple Prime & 74.99\% & 65\% & 0\% \\
        Broadcast Mode & 14.10\% & 0\% & 0\% \\
        \bottomrule
    \end{tabular*}
\end{table}

Fig.~\ref{fig:link-failure-robustness} shows that, for failure rates below about $70\%$, CayleyTopo maintains substantially lower $T90$ than ExpoComm and the other rule-based topologies; averaged over the $30\%$-$70\%$ regime, CayleyTopo is about $40\%$ faster than ExpoComm in terms of $T90$. At $85\%$ failures, all curves approach the same high delay ($\approx\!120$-$121$ rounds), i.e., performance becomes comparable to the broadcast-style cap, because the underlying graphs are heavily fragmented. Table~\ref{tab:robustness-85} reports the complementary LCC and Pr80 metrics at $85\%$ failures: CayleyTopo attains the strongest giant component and full Pr80-R/Pr80-D under both failure models, whereas distance-biased removal can yield Pr80-D$=0$ even when Pr80-R remains nonzero (e.g., ExpoComm).

\begin{table}[t]
    \centering
    \small
    \setlength{\tabcolsep}{4pt}
    \renewcommand{\arraystretch}{1.15}
    \caption{Per-step bandwidth consumption over $50$ simulation time steps under the hybrid communication schedule (same scenario as Fig.~\ref{fig:comm_count}). Values are scaled by $10^{-3}$ relative to raw simulator counts. Lower average indicates lower overhead; standard deviation and range (max-min) quantify temporal variability.}
    \label{tab:comm_bandwidth}
    \begin{tabular*}{\columnwidth}{@{\extracolsep{\fill}}@{}l r r r@{}}
        \toprule
        Topology & Avg.  & Std.\ dev.  & Range  \\
        \midrule
        CayleyTopo & $140.7$ & $16.1$ & $76.8$ \\
        ExpoComm & $325.3$ & $212.1$ & $614.4$ \\
        Fibonacci & $271.3$ & $188.7$ & $599.0$ \\
        Simple Prime & $178.5$ & $82.2$ & $399.4$ \\
        Broadcast Mode & $768.0$ & $0.0$ & $0.0$ \\
        \bottomrule
    \end{tabular*}
\end{table}
	
\begin{figure}[t]
    \centering
    \includegraphics[width=0.4\textwidth]{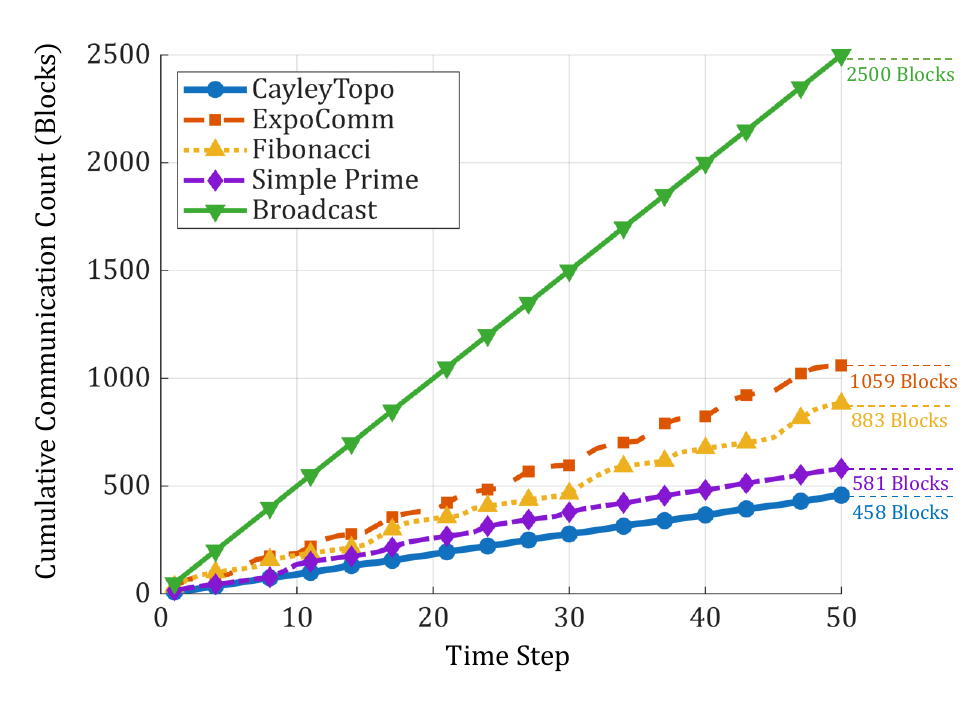}
    \caption{Cumulative communication-count over time. Lower values indicate fewer occupied transmission blocks.}
    \label{fig:comm_count}
\end{figure}

\subsection{Communication Overhead in Information Sharing}
Beyond latency and robustness, a topology must also support efficient communication under realistic message loads. We simulate a scenario where agents exchange information over $50$ discrete time steps using a hybrid communication schedule (messages are sent only when needed, as opposed to broadcast). We track the cumulative number of transmissions (i.e., resource-block occupancy) and the resulting bandwidth consumption.
Table~\ref{tab:comm_bandwidth} and Fig.~\ref{fig:comm_count} present communication-load results. The optimized topology consistently incurs the lowest cumulative transmissions and bandwidth usage among all compared designs; in particular, CayleyTopo achieves the lowest average per-step bandwidth and comparatively small temporal variability, whereas broadcast is non-fluctuating but incurs the highest average load. Notably, it achieves these savings while maintaining the fast dissemination and robustness demonstrated earlier. This confirms that CayleyTopo strikes a favorable balance between connectivity and communication overhead.

\subsection{Comparison to the Moore bound}
\begin{figure}[t]
    \centering
    \includegraphics[width=0.4\textwidth]{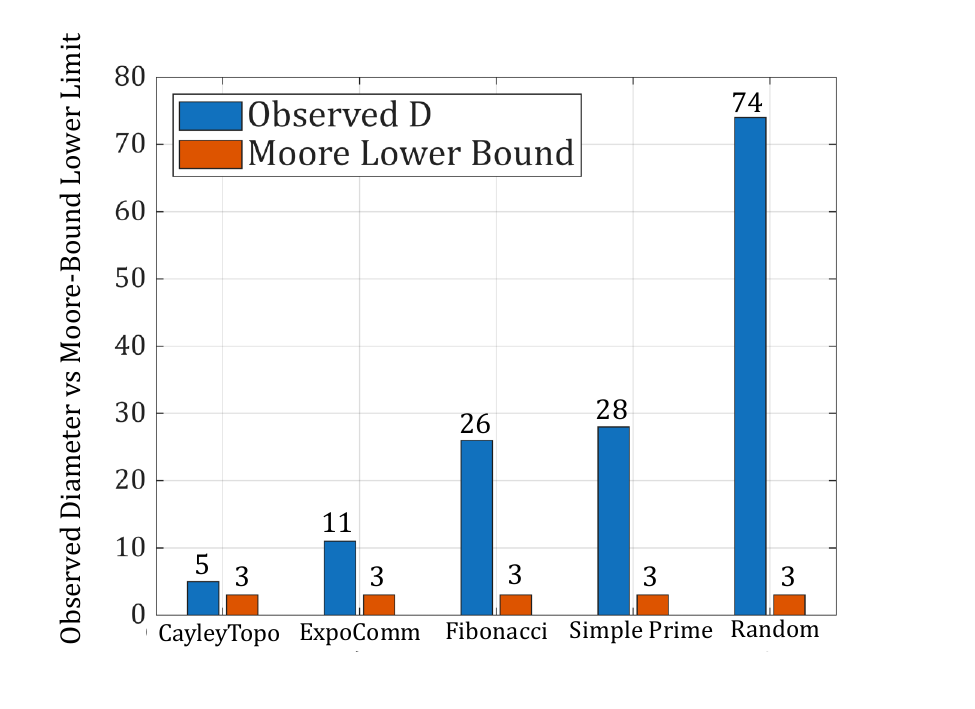}
    \caption{The graph diameter against the Moore bound under matched degree.}
    \label{fig:moore}
\end{figure}

To contextualize the optimized topology quality, we compare the graph diameter against the Moore bound under the same degree budget. Fig.~\ref{fig:moore} shows that CayleyTopo remains close to the theoretical limit while preserving stronger practical behavior in dissemination and communication load.

\begin{figure}[t]
    \centering
    \includegraphics[width=0.4\textwidth]{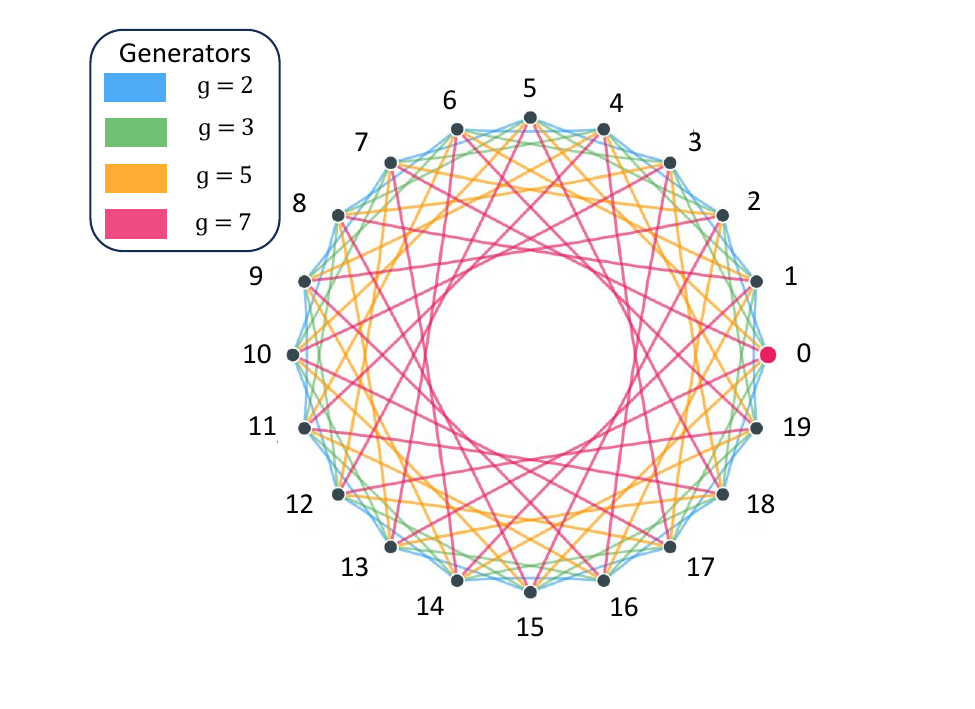}
    \caption{Visualization of CayleyTopo on a reduced ring, where $N=20$.}
    \label{fig:topology-vis}
\end{figure}

Finally, for an intuitive structural view, Fig.~\ref{fig:topology-vis} presents the CayleyTopo on a reduced-size ring with $N=20$ agents.

\section{Conclusion}
This work shifts the perspective on large-scale multi-agent communication from hand-crafted rules to optimizable design. By formulating topology selection as a discrete optimization over circulant Cayley graphs, we demonstrate that superior communication structures can be discovered rather than prescribed.
The resulting CayleyTopo family is not only performant but also practical: it remains vertex-transitive, deterministic, and easy to implement, requiring no per-agent coordination beyond the shared generator set. 

More broadly, our study suggests that the communication graph itself should be treated as a first-class design variable, not a fixed constraint. As multi-agent systems scale to thousands or millions of agents, we believe that the paradigm of CayleyTopo, grounded in algebraic graph theory, will become a cornerstone for scalable, robust, and efficient multi-agent communication.

\bibliographystyle{ieeetr}
\bibliography{ref}

@inproceedings{li2025exponential,
  title={Exponential Topology-enabled Scalable Communication in Multi-agent Reinforcement Learning},
  author={Li, Xinran and Wang, Xiaolu and Bai, Chenjia and Zhang, Jun},
  booktitle={The Thirteenth International Conference on Learning Representations (ICLR)},
  year={2025}
}

@article{schulman2017ppo,
  title={Proximal Policy Optimization Algorithms},
  author={Schulman, John and Wolski, Filip and Dhariwal, Prafulla and Radford, Alec and Klimov, Oleg},
  journal={arXiv preprint arXiv:1707.06347},
  year={2017}
}

@article{chen2019uav,
  title={When {UAV} swarm meets edge-cloud computing: The {QoS} perspective},
  author={Chen, Wuhui and Liu, Baichuan and Huang, Huawei and Guo, Song and Zheng, Zibin},
  journal={IEEE Network},
  volume={33},
  number={2},
  pages={36--43},
  year={2019}
}

@article{shao2021federated,
  title={Federated edge learning with misaligned over-the-air computation},
  author={Shao, Yulin and G{\"u}nd{\"u}z, Deniz and Liew, Soung Chang},
  journal={IEEE Transactions on Wireless Communications},
  volume={21},
  number={6},
  pages={3951--3964},
  year={2021}
}

@article{shao2024theory,
  title={A theory of semantic communication},
  author={Shao, Yulin and Cao, Qi and G{\"u}nd{\"u}z, Deniz},
  journal={IEEE Trans. Mobile Comp.},
  volume={23},
  number={12},
  pages={12211--12228},
  year={2024}
}

@article{schulman2015gae,
  title={High-Dimensional Continuous Control Using Generalized Advantage Estimation},
  author={Schulman, John and Moritz, Philipp and Levine, Sergey and Jordan, Michael and Abbeel, Pieter},
  journal={arXiv preprint arXiv:1506.02438},
  year={2015}
}

@article{olfati2007consensus,
  title={Consensus and Cooperation in Networked Multi-Agent Systems},
  author={Olfati-Saber, Reza and Fax, J. Alex and Murray, Richard M.},
  journal={Proceedings of the IEEE},
  volume={95},
  number={1},
  pages={215--233},
  year={2007},
  publisher={IEEE}
}

@article{sukhbaatar2016learning,
  title={Learning multiagent communication with backpropagation},
  author={Sukhbaatar, Sainbayar and Fergus, Rob and others},
  journal={Advances in neural information processing systems},
  volume={29},
  year={2016}
}

@article{foerster2016learning,
  title={Learning to communicate with deep multi-agent reinforcement learning},
  author={Foerster, Jakob and Assael, Ioannis Alexandros and De Freitas, Nando and Whiteson, Shimon},
  journal={Advances in neural information processing systems},
  volume={29},
  year={2016}
}

@inproceedings{das2019tarmac,
  title={Tarmac: Targeted multi-agent communication},
  author={Das, Abhishek and Gervet, Th{\'e}ophile and Romoff, Joshua and Batra, Dhruv and Parikh, Devi and Rabbat, Mike and Pineau, Joelle},
  booktitle={International Conference on machine learning},
  pages={1538--1546},
  year={2019},
  organization={PMLR}
}

@article{jiang2018learning,
  title={Learning attentional communication for multi-agent cooperation},
  author={Jiang, Jiechuan and Lu, Zongqing},
  journal={Advances in neural information processing systems},
  volume={31},
  year={2018}
}

@book{godsil2013algebraic,
  title={Algebraic graph theory},
  author={Godsil, Chris and Royle, Gordon F},
  year={2013},
  publisher={Springer Science \& Business Media}
}

@article{miller2012moore,
  title={Moore graphs and beyond: A survey of the degree/diameter problem},
  author={Miller, Mirka and Sir{\'a}n, Jozef},
  journal={The electronic journal of combinatorics},
  pages={DS14--May},
  year={2012}
}

@book{ireland1990classical,
  title={A classical introduction to modern number theory},
  author={Ireland, Kenneth and Rosen, Michael Ira},
  volume={84},
  year={1990},
  publisher={Springer Science \& Business Media}
}

@book{tao2006additive,
  title={Additive combinatorics},
  author={Tao, Terence and Vu, Van H},
  volume={105},
  year={2006},
  publisher={Cambridge University Press}
}
	
\end{document}